\documentclass[preprint,showpacs,amsmath,amssymb,nofootinbib]{revtex4}
\pdfoutput=1
\usepackage{rotating}
\usepackage{array}
\usepackage{epsfig}
\usepackage{graphicx}
\usepackage{amssymb}

\newcommand{\be}{\begin{e
\usepackage{latexsym}qnarray}}
\newcommand{\bb}{\bibitem}
\newcommand{\ee}{\end{eqnarray}}

\newcommand{\fig}{\begin{figure}}
\newcommand{\ef}{\end{figure}}
\newcommand{\bc}{\begin{center}}
\newcommand{\ec}{\end{center}}
\newcommand{\mb}{\mathbf}
\newcommand{\nm}{\mbox}

\newcommand{\bn}{\begin{enumerate}}
\newcommand{\en}{\end{enumerate}}

\newcommand{\bz}{\begin{itemize}}
\newcommand{\ez}{\end{itemize}}
\newcommand{\ct}{\centerline}
\newcommand{\ep}{\epsfig}
\newcommand{\cp}{\caption}

\newcommand{\ba}{\begin{array}} 
\newcommand{\ea}{\end{array}}

\newcommand{\bt}{\begin{tabular}}
\newcommand{\et}{\end{tabular}}

\newcommand{\ub}{\underbrace}

\newcommand{\mc}{\mathcal}

\newcommand{\bd}{\begin{displaymath}}
\newcommand{\ed}{\end{displaymath}}
\newcommand{\nn}{\nonumber}
\newcommand{\ben}{\begin{eqnarray*}}
\newcommand{\een}{\end{eqnarray*}}

\newcommand{\ga}{\gamma}

\newcommand{\bq}{\begin{quote}}
\newcommand{\eq}{\end{quote}}

\newcommand{\gsim}{\gtrsim}
\newcommand{\lsim}{\lesssim}

\begin{document}
\def\bea{\begin{eqnarray}}
\def\eea{\end{eqnarray}}
\def\nn{\nonumber}
\newcommand{\snu}{\tilde \nu}
\newcommand{\sll}{\tilde{l}}
\newcommand{\asnu}{\bar{\tilde \nu}}
\newcommand{\stau}{\tilde \tau}
\newcommand{\dmsnu}{{\mbox{$\Delta m_{\tilde \nu}$}}}
\newcommand{\mt}{{\mbox{$\tilde m$}}}

\renewcommand\epsilon{\varepsilon}
\def\be{\begin{eqnarray}}
\def\ee{\end{eqnarray}}
\def\lla{\left\langle}
\def\rra{\right\rangle}
\def\za{\alpha}
\def\zb{\beta}
\def\lsim{\mathrel{\raise.3ex\hbox{$<$\kern-.75em\lower1ex\hbox{$\sim$}}} }
\def\gsim{\mathrel{\raise.3ex\hbox{$>$\kern-.75em\lower1ex\hbox{$\sim$}}} }
\newcommand{\Rbs}{\mbox{${{\scriptstyle \not}{\scriptscriptstyle R}}$}}

\draft
\title{Constraining parameter space  in type-II two-Higgs doublet model \\ in light of  a 126 GeV Higgs boson}

\thispagestyle{empty}
\author{ H. S. Cheon \footnote{E-mail:
        hscheon@gmail.com}, Sin Kyu Kang \footnote{E-mail:
        skkang@snut.ac.kr}}
\affiliation{Institute of Convergence Fundamental Studies, Seoul National University of Science and Technology, Seoul 139-743, Korea }

\begin{abstract}
We explore the implications of a 126 GeV Higgs boson indicated by the recent LHC results
for  two-Higgs doublet model (2HDM).
Identifying the 126 GeV Higgs boson as either the lighter or heavier of  CP even neutral  Higgs bosons in 2HDM,
we examine how the masses of Higgs fields and mixing parameters can be constrained by  the theoretical conditions and experimental constraints. The theoretical conditions taken into account are the vacuum stability, perturbativity and unitarity required to
be satisfied up to a cut-off scale. We also show how bounds on the masses of Higgs bosons and mixing parameters depend on the cut-off scale.
In addition, we investigate whether the allowed regions of parameter space can accommodate particularly the enhanced di-photon signals, $ZZ^{\ast}$ and $WW^{\ast}$ decay modes of the Higgs boson, and examine the prediction of the signal strength of $Z\gamma$ decay mode  for the allowed regions of the parameter space.
\end{abstract}

\maketitle \thispagestyle{empty}
\section{Introduction}
Both the ATLAS and CMS experiments have  discovered a new particle consistent with the Higgs boson  \cite{higgs} with a  mass of  around 126 GeV at about $ 5 \sigma$ significance \cite{ATLAS, CMS}.
A common belief among particle physicists that  the SM is not  the ultimate
theory of fundamental interactions calls for new physics  beyond the SM, such as supersymmetry (SUSY) and extra dimension models.
Many new physics
beyond the SM contain more than one Higgs doublet  of the SM \cite{Branco}.
In this regards, it must deserve to examine  whether  signals detected at the LHC  imply the existence of more Higgs sectors or not.

The purpose of this work is to examine the implications of a 126 GeV Higgs boson indicated by the recent LHC results
for  two-Higgs doublet model (2HDM).
We will focus on how severe  the theoretical conditions and experimental results on the Higgs sectors
can constrain  the masses of Higgs fields and mixing parameters in 2HDM in the light of a 126 GeV Higgs boson.
The theoretical conditions taken into account are the vacuum stability, perturbativity and unitarity which are required to be satisfied up to a cut-off scale.
Then one can obtain constraints on the couplings of the Higgs potential
in 2HDM, which in turn lead to bounds on the masses of scalar bosons as well as mixing parameters.
Although there are a few works on the estimation of  bounds on the masses of scalar fields in 2HDM by applying the vacuum stability,
perturbativity \cite{2HDMvacuum, Ferreira} and  unitarity  \cite{unitarity},
our new points  are to show
how the parameter spaces in 2HDM are constrained by those theoretical conditions applied up to a cut-off scale
by identifying the 126 GeV Higgs boson as either  lighter or heavier of CP even neutral  scalar bosons, and to see how bounds on the masses of scalar bosons depend on the cut-off scale.
In addition, we will examine how experimental constraints on the parameters of scalar bosons from the LEP can constrain the parameter spaces further. As expected, LEP results can severely constrain the parameter space for scalar bosons in the scenario that
the new scalar boson observed at the LHC is the heavier CP even neutral scalar boson in the 2HDM.
Finally, we will investigate whether the allowed regions of parameter space can accommodate the enhanced di-photon signals, $ZZ^{\ast}$
 and $WW^{\ast}$ decay modes of the Higgs boson observed at the LHC,
and examine the prediction of the signal strength of $Z\gamma$ decay mode for the allowed parameter regions.

\section{Higgs sector in 2HDM, theoretical and experimental constraints}
The renormalizable gauge  invariant  scalar potential of 2HDM with softly broken $Z_2$ symmetry we consider  is given by \cite{2HDM}
\be V= && m^2 _{11} \Phi^\dagger _1 \Phi_1 + m^2 _{22} \Phi^\dagger _2 \Phi_2 - (m^2 _{12} \Phi^\dagger _1 \Phi_2 + \nm{h.c.}) \nn \\
&& + \frac{1}{2}\lambda_1 (\Phi^\dagger _1 \Phi_1)^2 + \frac{1}{2}\lambda_2 (\Phi^\dagger _2 \Phi_2 )^2 + \lambda_3 (\Phi^\dagger _1 \Phi_1) (\Phi^\dagger _2 \Phi_2) \nonumber \\
&& + \lambda_4 (\Phi^\dagger_1 \Phi_2 ) (\Phi^\dagger _2 \Phi_1)  + \Big \{ \frac{1}{2} \lambda_5 (\Phi^\dagger _1 \Phi_2 )^2
+ \nm{h.c.} \Big \},  \label{2hd pot}
\ee
where $\Phi_1$ and $\Phi_2$ are two complex $SU(2)_L$ Higgs doublet  fields with  $Y=1$.
We note that the dangerous FCNC does not occur in the form of the scalar potential given by Eq.(\ref{2hd pot}) even if  non-zero $m^2_{12}$ softly breaking the $Z_2$ symmetry is allowed.
Depending on how to couple the Higgs doublets to the fermions, 2HDMs are classified into four types \cite{Branco}. Among them, the Yukawa couplings of type II 2HDM arises in the minimal supersymmetric standard model which is one of the most promising candidates for the new physics model beyond the SM.
In this paper, we focus on the type-II 2HDM, in which the one Higgs doublet $\Phi_1$ couples only to the down type quarks and the charged leptons while the another Higgs doublet $\Phi_2$ couples only to the up-type quarks.
 We require that the scalar potential  conserves the CP symmetry, which is achieved by taking all the parameters in Eq.(\ref{2hd pot}) to be real and the squared mass of pseudo-scalar $m^2 _A$ to be greater than  $ |\lambda_5| v^2$ for the absence of explicit  and  spontaneous CP violation, respectively \cite{2HDM}.
After spontaneous symmetry breaking, the Higgs doublets have the vacuum expectation values as follows,
\be <\Phi_1> = \frac{1}{\sqrt{2}} \left ( \ba{c} 0 \\ v_1 \ea \right ), \qquad <\Phi_2> = \frac{1}{\sqrt{2}} \left ( \ba{c} 0 \\ v_2 \ea \right ),
\ee
where
$ v^2 \equiv v^2 _1 + v^2 _2 =  (246 ~\mbox{GeV})^2$ and $v_2/v_1 = \tan\beta .$
We take  $v_1$ and $v_2$ to be positive, so that $0\leq \beta \leq \pi/2$ is allowed.
There are five physical Higgs particles in 2HDMs : two CP-even Higgs $h$ and $H$ ($M_h \leq M_H$), a CP-odd Higgs $A$ and a charged Higgs pair ($H^\pm$).
Following \cite{2HDM},
the squared masses for the CP-odd and charged Higgs states are given by
\be M^2 _A = \frac{m^2 _{12}}{s_\beta c_\beta} -\lambda_5 v^2 ,  \quad
M^2 _{H^\pm} =  M^2 _A + \frac{1}{2} v^2 (\lambda_5 - \lambda_4),
\label{cp odd charged masses}
\ee
and the squared masses for  neutral Higgs ($M_H \geq M_h$) are given by
\be
M^2 _{H, h} = \frac{1}{2}\Big  [  A+B \pm \sqrt{(A - B )^2 + 4 C^2 }~ \Big ], \label{cp even mass 2}
\ee
where
$A=\lambda_1 v^2_1+m^2_{12} t_\beta, B=\lambda_2 v^2_2+m^2 _{12}/ t_\beta$ and $C=(\lambda_3+\lambda_4+\lambda_5)v_1v_2 -m^2 _{12}$ with $s_\beta = \sin\beta$, $c_\beta = \cos\beta$, and $t_\beta=\tan\beta$.
The couplings of the two neutral CP even Higgs bosons to fermions and bosons relative to the SM couplings in type II 2HDM are shown in Table \ref{tab1}.
\begin{table}
\caption{\label{tab1} Neutral Higgs couplings relative to the SM couplings in Type II 2HDM.
$D,L,U,W,Z$ and $A$ stand for down-type quarks, charged leptons, up-type quarks, two weak gauge bosons and CP odd Higgs, respectively.}

\begin{tabular}{|c|c|c|}
\hline \hline
& Light Higgs ($h$) & Heavy Higgs ($H$) \\
\hline
D,L &  $- \frac{\sin\alpha}{\cos\beta}$ & $\frac{\cos\alpha}{\cos\beta}$ \\
U & $\frac{\cos\alpha}{\sin\beta}$ & $\frac{\sin\alpha}{\sin\beta}$ \\
W or Z & $\sin(\beta -\alpha)$ & $\cos(\beta -\alpha)$ \\
AZ & $- \cos(\beta -\alpha)$ & $-\sin(\beta -\alpha)$ \\
\hline \hline
\end{tabular}

\end{table}

The stable  vacuum guaranteed when the scalar potential (\ref{2hd pot}) is bounded from below can be obtained only if  the following conditions are satisfied \cite{Ferreira, 2HDM,2hdz2,Ivanov}
\be && \lambda_{1,2} > 0, \quad  \lambda_3 > -\sqrt{\lambda_1 \lambda_2}, \quad  \lambda_3 + \lambda_4 - |\lambda_5| > - \sqrt{\lambda_1 \lambda_2}.  \label{stability}
\ee
Since radiative corrections give rise to the modification of the couplings in the scalar potential,
we need to require that  the stability conditions (\ref{stability}) are valid for all energy scales up to a cut-off scale $\Lambda$.
As is known,  the stability conditions (\ref{stability}) can lead us to lower bounds on
the couplings $\lambda_i$ \cite{2HDM}, which in turn give rise to bounds on the masses of the
Higgs fields.
In addition,  we require the perturbativity  for the quartic couplings $\lambda_i $ in the scalar potential at all scales up to the cut-off scale $\Lambda$ and unitarity at the cut-off scale \cite{unitarity}.
It is worthwhile to notice that those theoretical conditions can constrain not only Higgs masses but also
mixing parameters $\tan\beta$ and $\alpha$ via the renormalization group (RG) evolutions.
In our numerical analysis, we used RG equations for the parameters
$m_{ii}^2, \lambda_{i}$ , gauge couplings $g_{i}$ and Yukawa couplings presented in ref.\cite{RGrunning}.
In particular, we take the top quark pole mass and QCD coupling constant at Z boson mass scale ($\alpha_s(M_Z)$)  to be 172 GeV and 0.1185,
respectively.

On the other hand, experimental results from the LEP give rise to constraints on the masses of Higgs bosons  and  the mixing parameters in the case that the masses of  light neutral Higgs bosons lie between 10 GeV and 150 GeV  \cite{LEP,LEP2} .
For the charged Higgs bosons, the experimental lower bound on their masses is $79.3$ GeV \cite{ALEPH}.
 The non-observation of $Z\rightarrow h A$ in the LEP experiment
indicates that  only the  Higgs masses satisfied with $M_h + M_A > M_Z$ are kinematically allowed \cite{zdecay}.
In addition,  when $M_h \lesssim 115$ GeV,
non-observation of the Higgsstrahlung process $e^+ e^- \rightarrow hZ \rightarrow b\bar{b}Z$  at the LEP constrains the parameter space of $\sin^2(\beta-\alpha)\times Br(h\rightarrow b\bar{b})$ and $M_h$\footnotemark
\addtocounter{footnote}{0}
\footnotetext{The parameter $\xi^2$ introduced in \cite{LEP} is equivalent to
$\sin^2(\beta - \alpha) $ in our model.}.
We also consider  the Higgs pair production process, $e^+ e^- \rightarrow hA  \rightarrow b\bar{b}b\bar{b}$,
if  they are kinematically allowed.
Non-observation of those Higgs pair productions can lead to the constraints on light neutral Higgs masses and mixing parameters
as shown in \cite{LEP2}.

In addition, we take into account the new physics contributions to the electroweak precision parameters $\rho_0$ and $S$, which are defined by \cite{pdg,Toussaint,kanemura,Baak,Neil}
\be \rho_0 &\equiv& \frac{M^2_W}{\rho M^2 _Z \cos^2 \theta_w} =1+\Delta \rho^{\rm new} _0,  \\
S &=& - \frac{1}{4\pi} [ F(M_{H^\pm}, M_{H^\pm})) - \sin^2 (\beta -\alpha) F (M_H, M_A) \nn \\
&& - \cos^2 (\beta -\alpha) F(M_h . M_A) ],  \label{S parameter}
\ee
where $\Delta \rho^{\rm new} _0 = \Delta \rho^{\rm 2HDM} - \Delta \rho^{\rm SM}$ and the formulae for $\Delta \rho^{\rm 2HDM}$ as well as $\Delta \rho^{\rm SM}$ are given in
 \cite{hunter, cheung, Chankowski}, and the function $F$ is given by \cite{Toussaint,kanemura,Baak}
\be F(x,y)  = - \frac{1}{3} \Big [\frac{4}{3} - \frac{x^2 \ln x^2 - y^2 \ln y^2}{x^2 - y^2}
 - \frac{x^2 + y^2}{(x^2 - y^2)^2} \Big ( 1+\frac{x^2 +y^2}{2} - \frac{x^2 y^2}{x^2 - y^2}\ln\frac{x^2}{y^2}\Big ) \Big ],
\ee
By fixing $U=0$, the allowed values of $\Delta \rho^{\rm new} _0$ and $S$ for $115.5~ \mbox{GeV} < M_{\rm SM ~higgs} < 127~ \mbox{GeV}$ are given by \cite{pdg}
\be - 0.0001 \leq &\Delta \rho^{\rm new} _0&  \leq 0.0012 \nn \\
-0.05 \leq &S& \leq 0.13. \label{ew-parameter}
\ee
We impose the conditions Eq.(\ref{ew-parameter}) in our numerical analysis.
On top of the constraints from $\Delta \rho^{\rm new}_0$ and $S$, we consider  the measurement of  $R_b \equiv \Gamma (Z \rightarrow b \bar{b})/\Gamma(Z \rightarrow \mbox{hadrons})$ \cite{pdg} as well as the experimental results of the process $b\rightarrow s \gamma$ \cite{bsg}, which give rise to the constraints on the $M_{H^\pm} - \tan \beta$ plane. In the Type-II 2HDM,  it is known that $R_b$ yields the strictest bound on the $M_{H^\pm} - \tan \beta$ plane in the small $\tan\beta$ region  \cite{rb,d2hdm}. The measurements of $B-\bar{B}$ mixing also lead to the constraints on  the $M_{H^\pm} - \tan\beta$ plane but  less severe ones in comparison with that from $R_b$ \cite{d2hdm}.
Combining the theoretical constraints with the experimental ones, we investigate how  the masses of Higgs bosons and mixing parameters can be constrained.

\section{Allowed regions of parameter spaces}
Let us study the implication of  the 126 GeV Higgs boson indicated by the recent LHC results by identifying  it as
the lighter or heavier of the CP even neutral Higgs bosons.
Instead of fixing a particular value of the Higgs boson mass, we broaden it to be  $124.9
~\mbox{GeV} \leq M_H(M_h) \leq 126.4~\mbox{GeV}$ and then investigate parameter space in consistent with the range.
In our numerical analysis, the scanned regions of the parameters, $\tan\beta$ and $m_{12}$, are
\begin{eqnarray}
0.3 \mb{\leq} \tan\beta \mb{\leq} 50,~~~0< m_{12} < 109~ (1000)~\mbox{GeV},~~\mbox{for}~~ 124.9 \leq M_H (M_h) \leq 126.4  \label{para}
\end{eqnarray}
Note that small $\tan\beta$ below 0.3 is ruled out  by breaking down of perturbativity of Higgs-top Yukawa coupling \cite{H-top}.
We observed from our numerical analysis that vacuum stability excludes the region of  $m_{12}~\geq~109$ GeV for the case of $M_H \sim 126$ GeV and $M_h \leq 100$ GeV.
For the case of $M_h~\sim~126$ GeV, there are allowed regions of the parameter space above $m_{12} = 1000$ GeV, but such large values of $m_{12}$ lead to large values of scalar masses in the 2HDM leading to so-called decoupling limit, so we cut the size of $m_{12}$ by 1000 GeV in our analysis.
Instead of getting the regions of the parameter space with solid boundaries, we plot allowed data points by randomly scanning
the input parameters such as $\lambda_{i(=1-5)},~ m_{12}$ and $\tan\beta$ restricted by Eq.(\ref{para}) and then picking out the data points satisfying theoretical conditions and experimental constraints.
\fig [h] \ct{\ep{figure=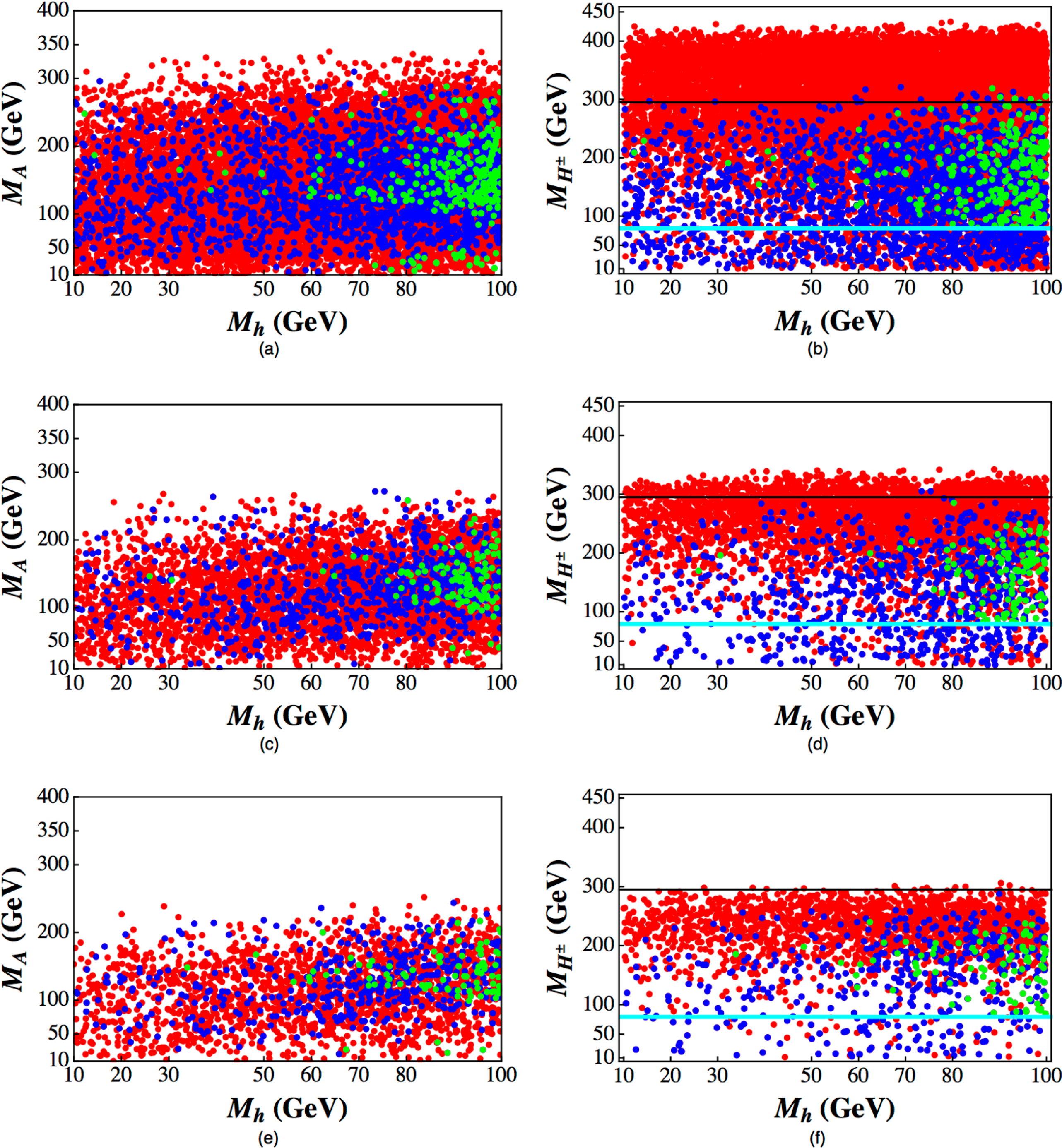, scale=0.51}}\cp{ Allowed regions  in the plains ($M_A, M_h)$ (left panels) and ($M_{H^\pm}, M_h$) (right panels) . The panels in the upper, middle and lower rows correspond to $\Lambda = 1, 14, 100$ TeV, respectively.
The territories covered by the red points are allowed by theoretical constraints. The  blue points survive the constraints from $\Delta \rho^{\rm new}_0, S$ and $R_b$. Further imposing LEP constraints, the green points finally survive.
 The black  horizontal lines correspond to  $b \rightarrow s \ga$ constraint \cite{btosgam} ($M_{H^{\pm}}=295$ GeV), and the cyan ones to the experimental lower bound on $M_H^{\pm}$ .}\label{fig1}
\ef
\subsection{Case for $M_H \sim 126$ GeV } \label{HA}
 Assuming that the mass of the heavier neutral CP even Higgs is around 126 GeV, let us examine how
the parameter space of Higgs masses and mixing parameters  can be constrained by  theoretical conditions  and experimental constraints explained in Sec. II. Also, we investigate how  the allowed regions of the parameter space depend on the cut-off scale.

Fig. \ref{fig1} shows how the regions of parameter spaces in the plains $(M_h, M_{A})$ (left-hand panels) and $(M_h, M_{H^{\pm}})$ (right-hand panels)  are constrained by the theoretical conditions and experimental results.
The panels in the upper, middle and lower rows correspond to the cases of the cut off scale $\Lambda \simeq 1,~14~$ and $100$ TeV, respectively.
The territories covered by the red points present the allowed regions by the theoretical conditions.
The blue regions survive the constraints on  $\Delta \rho^{\rm new}_0$,  parameter $S$ and $R_b$.
Further imposing constraints coming from the direct searches for Higgs fields via the Higgsstrahlung and Higgs pair productions at the LEP and the bound on $M_{H^{\pm}}$ from ALEPH, the green data points finally survive.
We see that the allowed regions get wider as the cut off scale gets lower.
The black horizontal lines correspond to the lower limit of $M_{H^{\pm}}$ coming from the experimental constraint
from $b\rightarrow s \gamma $ \cite{btosgam}, and thus the regions below the lines are excluded if no new effects on flavor physics are introduced in 2HDM.
We also display the cyan lines corresponding to the lower limit of the charged higgs mass from the ALEPH experiment, $M_{H^\pm} = 79.3$ GeV \cite{ALEPH}.
From our numerical analysis, we found that the constraint from $b\rightarrow s \gamma$ excludes all parameter regions survived other constraints for $\Lambda\gsim14 $ TeV
in the case of $M_H \sim~126$ GeV.

\fig [h] \ct{\ep{figure=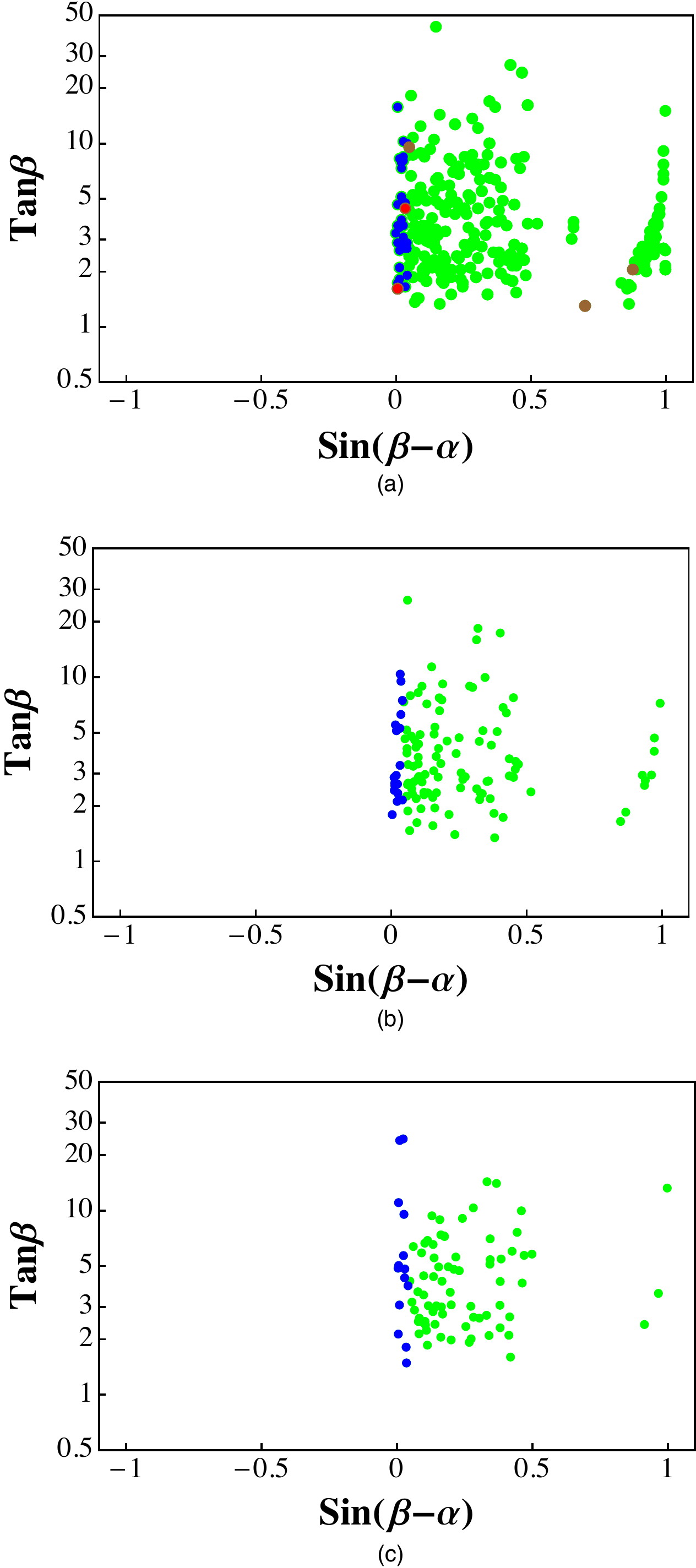, scale=0.55}} \cp{ Plots of ~ $\sin(\beta-\alpha) ~ {\it vs.}~\tan\beta $~ allowed by
theoretical conditions and experimental constraints for  $\Lambda=$1 (a),~~14 (b),~~100 (c) TeV, respectively.
The brown and red points survive all the constraints we consider, whereas the green and blue points
survive all the constraints except for $b\rightarrow s \gamma$ and thus correspond to $M_{H^{\pm}}\leq 295 $ GeV.  Points consistent with SM-like Higgs are displayed in blue and red points.  }\label{fig2}
\ef

In Fig. \ref{fig2}, the points represent the parameter space in the plain  $(\sin(\beta-\alpha), ~\tan\beta)$  constrained by the theoretical conditions and experimental constraints for  $\Lambda=$1 (a),~14 (b),~100 (c) TeV, respectively.
The brown and red points survive all the constraints we consider, whereas  the green and blue points survive all the constraints except for $b\rightarrow s \gamma$ and thus correspond to $M_{H^{\pm}}\leq 295 $ GeV.
In particular, we display the points consistent with SM-like Higgs, $\cos(\beta -\alpha) \sim 1$, in  blue  ($M_{H^{\pm}}\leq 295$ GeV) and  red  ($M_{H^{\pm}}> 295$ GeV).
As explained  above, we do not see any data points for $\Lambda \gsim 14$  TeV survived the constraint from $b\rightarrow s \gamma$.
We see that the region of $\tan\beta < 1$  is excluded in all cases we consider.
It is worthwhile to notice that  the mixing parameter $\beta-\alpha$ can be constrained by not only the LEP experiments but also
 the theoretical conditions.
As the cut off scale increases, the allowed regions get narrowed as shown in Fig. \ref{fig2}.

\fig [h] \ct{\ep{figure=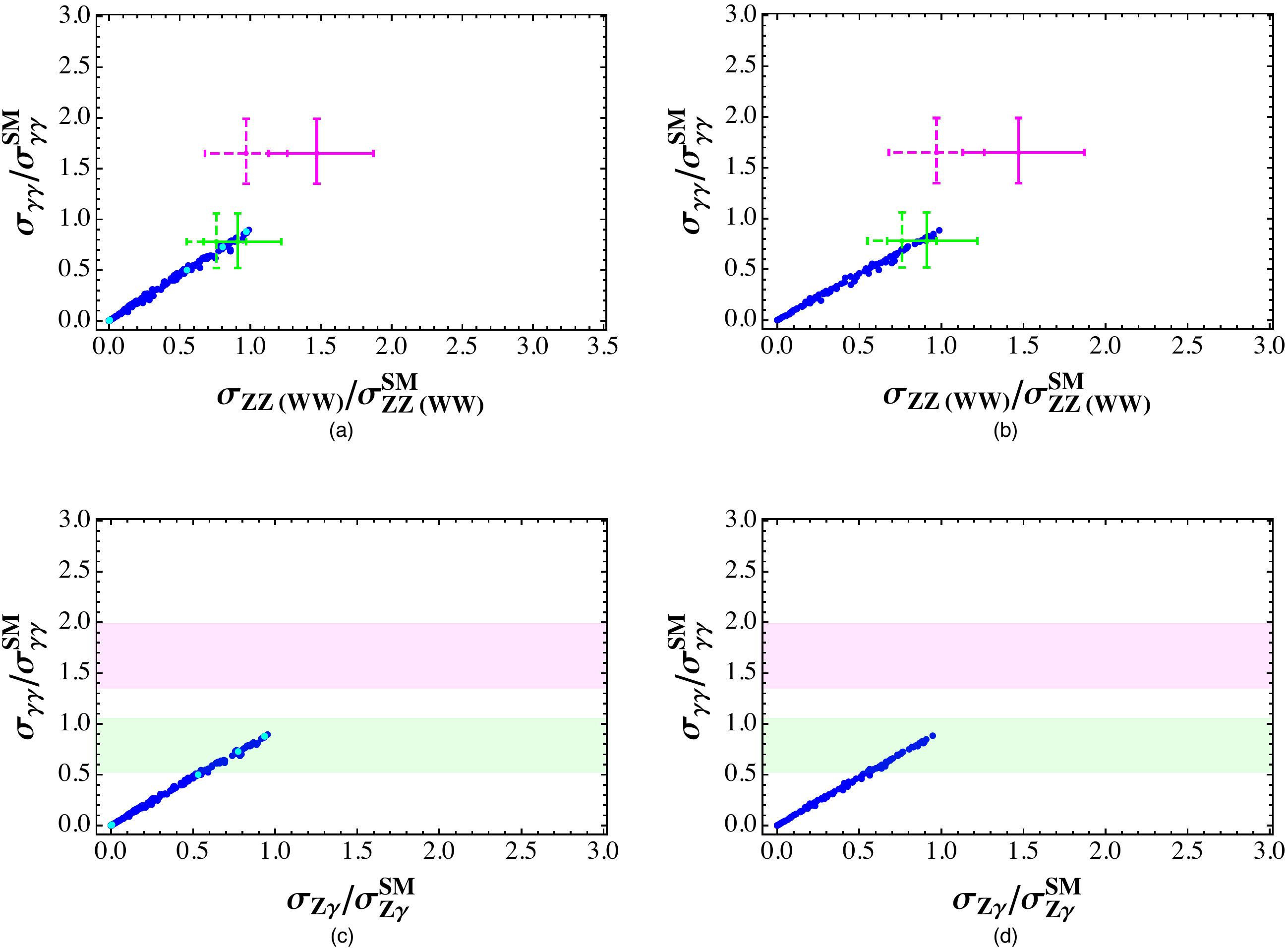, scale=0.49}}
\cp{ Predictions of   $\sigma_{\ga \ga}/\sigma^{SM}_{\ga \ga} ~ {\it vs.}~
\sigma_{ZZ^{\ast}(WW^{\ast})}/\sigma_{ZZ^{\ast}(WW^{\ast})}^{SM} $ (a,b) and $\sigma_{\ga \ga}/\sigma^{SM}_{\ga \ga} ~ {\it vs.}~\sigma_{Z\ga}/\sigma_{Z\ga}^{SM}$ (c,d) for
$\Lambda=1 $ (left) and 14 (right) TeV.
All blue points correspond to the green ones  in Figs. \ref{fig1} and \ref{fig2}.
The magenta (green) cross-bars represent  the ATLAS (CMS) results.
The dashed (solid) cross-bars correspond to the experimental results for $WW^\ast$ ($ZZ^\ast$) channel  given by Eq. \ref{WW} (\ref{ZZ}), and the  magenta (green) horizontal shaded region  to the di-photon channel given by Eq. \ref{gam ATLAS} (\ref{gam CMS}). }\label{fig3}
 \ef

\fig [h] \ct{\ep{figure=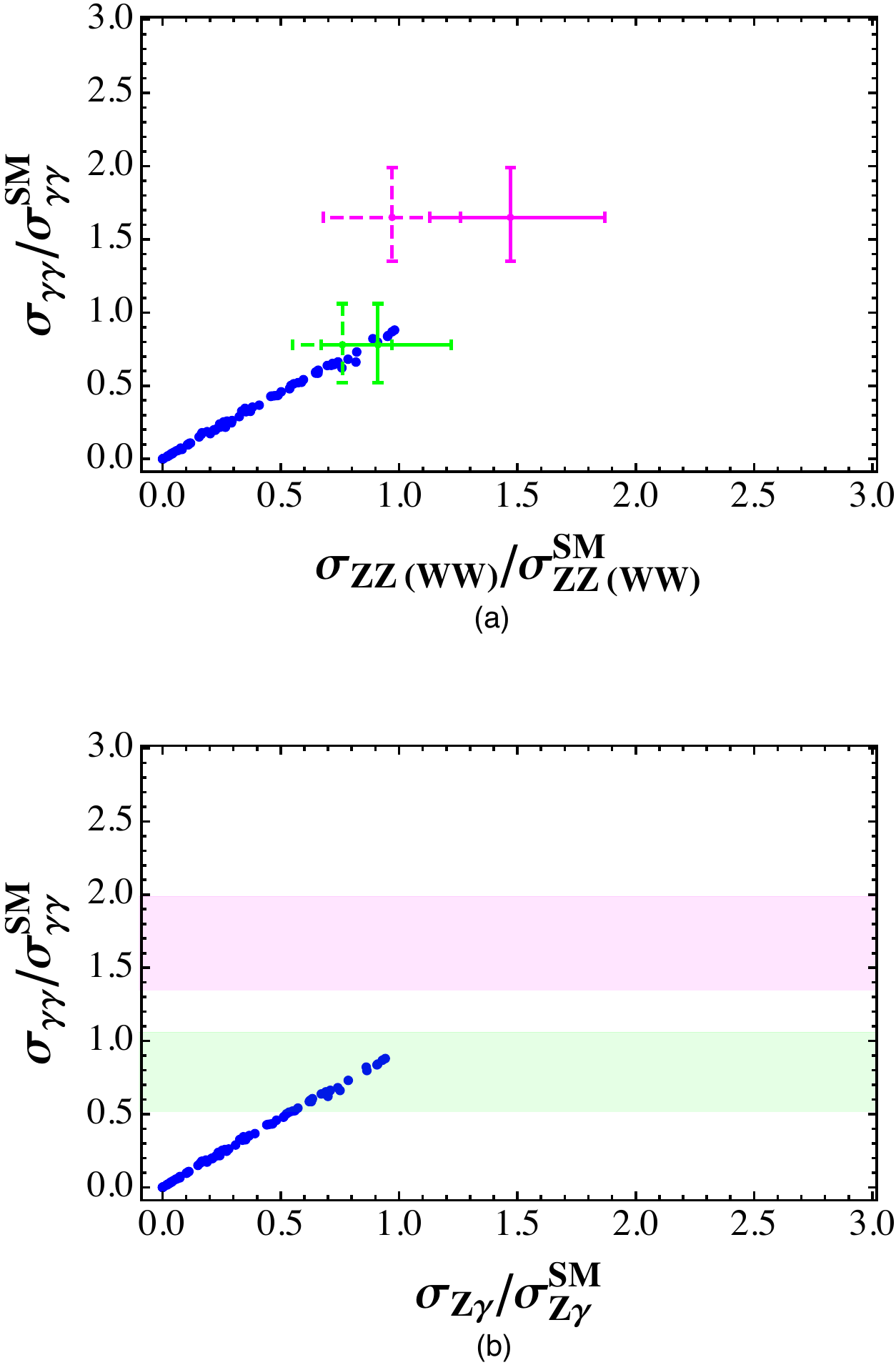, scale=0.49}} \cp{The same as Fig. \ref{fig3}, but $\Lambda =100$ TeV.}\label{fig3-2}
\ef

Let us discuss how the allowed regions obtained above can be confronted with the recent LHC data by considering the channels $H\rightarrow \ga\ga$, $H\rightarrow W W^* \rightarrow l\nu l \nu$, $H \rightarrow Z Z^* \rightarrow 4 l $ directly searched to probe the SM-like Higgs boson at the LHC.
The recent experimental results of the signal strengths for  $WW^{\ast}$ and $ZZ^{\ast}$ decay modes are given by
\cite{ATLAS,CMS},
\begin{eqnarray}
 \sigma_{WW^{\ast}}/\sigma_{WW^{\ast}}^{SM} &=& \left \{\begin{array}{lll}
                                                    0.97 \pm 0.29 && (\mbox{ATLAS}), \label{ww ATLAS} \\
                                                    0.76 \pm 0.21 && (\mbox{CMS}) , \label{ww CMS}
                                                     \end{array}\right.  \label{WW} \\
 \sigma_{ZZ^{\ast}}/\sigma_{ZZ^{\ast}}^{SM} &=& \left \{ \begin{array}{llll}
                                                     1.47^{+0.4} _{- 0.34} &&& (\mbox{ATLAS}) , \label{zz ATLAS}\\
                                                     0.91^{+0.31} _{-0.24} &&& (\mbox{CMS}) , \label{zz CMS}
                                                      \end{array}  \label{ZZ}\right.
\end{eqnarray}
where $\sigma_{VV^{\ast}} = \sigma(h)_{\rm prod} \times Br(h \rightarrow VV^{\ast})$ with $V=(W,Z)$.
The results are not incompatible with the SM predictions.
As for the measurements for the di-photon channel,  the current ATLAS results show a deviation from the SM prediction \cite{ATLAS}
\be \sigma_{\ga\ga}/\sigma_{\ga\ga}^{SM}= 1.65^{+0.34} _{-0.30}, ~~~\label{gam ATLAS}\ee  
whereas the CMS results appears to be compatible with the SM prediction \cite{CMS}
\be  \sigma_{\ga\ga}/\sigma_{\ga\ga}^{SM} = 0.78^{+0.28} _{-0.26}. \label{gam CMS} \ee

In Fig. \ref{fig3}-(a,b), we display  plots of  $\sigma_{\ga \ga}/\sigma^{SM}_{\ga \ga}$
~{\it vs.} ~$\sigma_{WW^{\ast}(ZZ^{\ast})}/\sigma^{SM}_{WW^{\ast}(ZZ^{\ast})}$  for the allowed regions of parameter space shown in Figs. \ref{fig1} and \ref{fig2}.  The left- and right-hand panels correspond to $\Lambda=1$ and $14$ TeV, respectively.
The magenta (green) cross-bars represent the ATLAS (CMS) experimental results.
The dashed (solid) cross-bars correspond to the experimental results for $WW^{\ast} (ZZ^{\ast})$ signal strengths.
In Fig. \ref{fig3}-(c,d), we present  the predictions of   $\sigma_{\ga\ga}/\sigma_{\ga\ga}^{\rm SM}
~{\it vs.} ~\sigma_{Z \gamma}/\sigma^{\rm SM}_{Z \gamma}$ for the same parameter
regions taken in Fig. \ref{fig3}-(a,b).
The magenta (green) shaded regions stand for the ATLAS (CMS) results for the di-photon signals.
Fig. \ref{fig3-2} shows the same as Fig. \ref{fig3} but for $\Lambda=100$ TeV.
All blue points in Figs. \ref{fig3} and \ref{fig3-2} correspond to the green ones  in Figs. \ref{fig1} and \ref{fig2}.
In particular,  we display in Fig. \ref{fig3}  the data points survived the constraint from $b\rightarrow s \gamma$ in cyan.
We see from Figs. \ref{fig3} and \ref{fig3-2} that there is no data point compatible with the enhanced di-photon signal measured at ATLAS, whereas there are parameter regions accommodating both the di-photon and vector boson pair signals observed from CMS.
To see in detail why the allowed regions of parameter space in this case can not lead to enhancement of di-phton signal, let us consider the formula of enhanced di-photon signal strength given by
\be \sigma_{\ga\ga}/\sigma^{\rm SM} _{\ga\ga} = \ub{\frac{\Gamma (gg \rightarrow \mc{H})}{ \Gamma (gg \rightarrow \mc{H} )_{\rm SM}}}_{(A)} \times \ub{\frac{ \Gamma (\mc{H} \rightarrow \ga\ga)}{ \Gamma (\mc{H} \rightarrow \ga\ga)_{\rm SM} }}_{(B)} \times \ub{\frac{ \Gamma^{\rm tot} (\mc{H}) _{\rm SM}}{ \Gamma^{\rm tot} (\mc{H})}}_{(C)},
\label{diphotons}
\ee
 where $\mc{H}$ is the heavy CP-even neutral higgs ($H$) or the light neutral higgs ($h$), and $\Gamma^{\rm tot} (\mc{H})$ denotes the total decay width of $\mc{H}$.
Because of the convention, $\sin(\beta -\alpha) \geq 0$, the coupling of heavy neutral higgs ($H$) to up-type quarks  relative to that of SM, $\frac{\sin\alpha}{\sin\beta}$, is smaller than one, whereas the coupling of $H$ to down type quarks (or charged leptons), $\frac{\cos\alpha}{\cos\beta}$, is larger than one. This indicates that both (A) and (C) in Eq. (\ref{diphotons}) should be smaller than one because the dominant contribution of gluon fusion is mediated by top quark loop, and $H \rightarrow b\bar{b}$ yields the most dominant contribution to the branching ratio of the 126 GeV $H$ decay. In addition,  the dominant contribution (mediated by $W$-loop) to the term (B) is proportional to $\cos(\beta-\alpha)$ which can not be larger than one.
Thus, the predictions of the signal strength for the di-photon channel can not be enhanced in this case.
%
\subsection{Case for $M_h=126$ GeV } \label{hb}
%
%

Assuming the lighter  CP-even neutral Higgs mass ($M_h$) is around 126 GeV, let us examine
how the parameter space of Higgs masses and mixing parameters can be constrained, and how constraining  the parameter spaces depends on the cut-off scale.
In this case, the experimental results coming from the direct search for the Higgs bosons at the LEP do not
further constrain the parameter space survived  the theoretical constraints.
%
\fig [h] \ct{\ep{figure=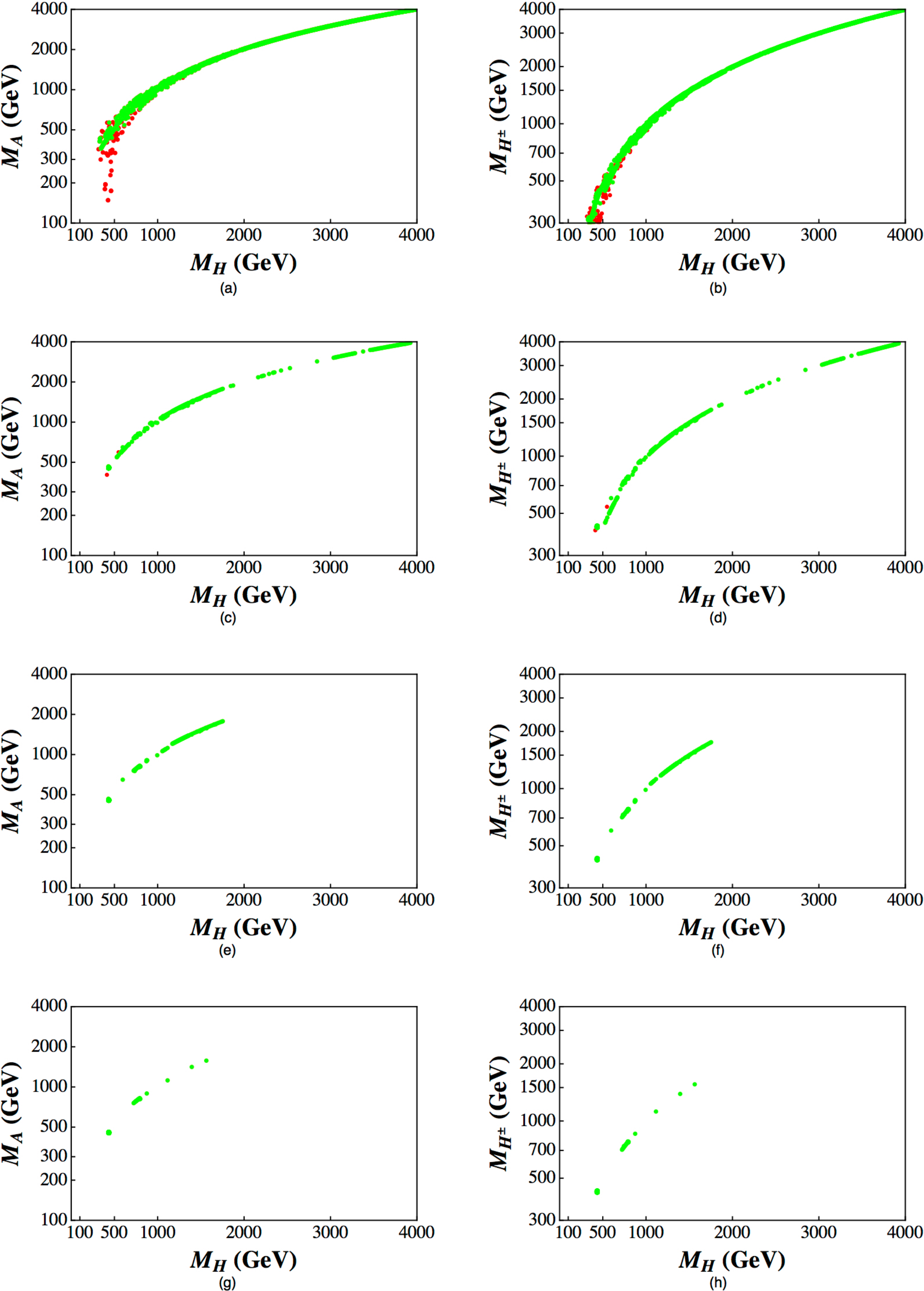,scale=0.44}}\cp{Allowed regions  in the plains ($M_{A},M_H)$ (left panels) and $(M_{H^\pm},M_H)$ (right panels) . The panels from top to bottom correspond to $\Lambda$ = 1, 10, 40 TeV and 100 TeV, respectively.
The red points are allowed only by stability, perturbativity and unitarity, and  the green ones survive all the constraints we consider.}\label{fig4}
\ef

In Fig. \ref{fig4}, we show how the regions of parameter spaces in the plains $(M_h, M_{A})$ (left panels) and $(M_h, M_{H^{\pm}})$ (right panels)  are constrained by the theoretical conditions and experimental results.
The panels from top to bottom correspond to the cases of the cut off scale $\Lambda \simeq$ 1 TeV, 10 TeV, 40 TeV and 100 TeV, respectively.
The territories covered by all the points present the allowed regions by stability, perturbativity and unitarity.
The green points survive the constraint on $\Delta \rho^{\rm new}_0$, $S$ and $R_b$.
In this case,  we scan only the parameter space satisfying the experimental constraint from $b\rightarrow s \gamma ~ (M_{H^{\pm}}>295$ GeV).
The allowed regions appear to get narrowed as the cut-off scale increases.

\fig [h] \ct{\ep{figure=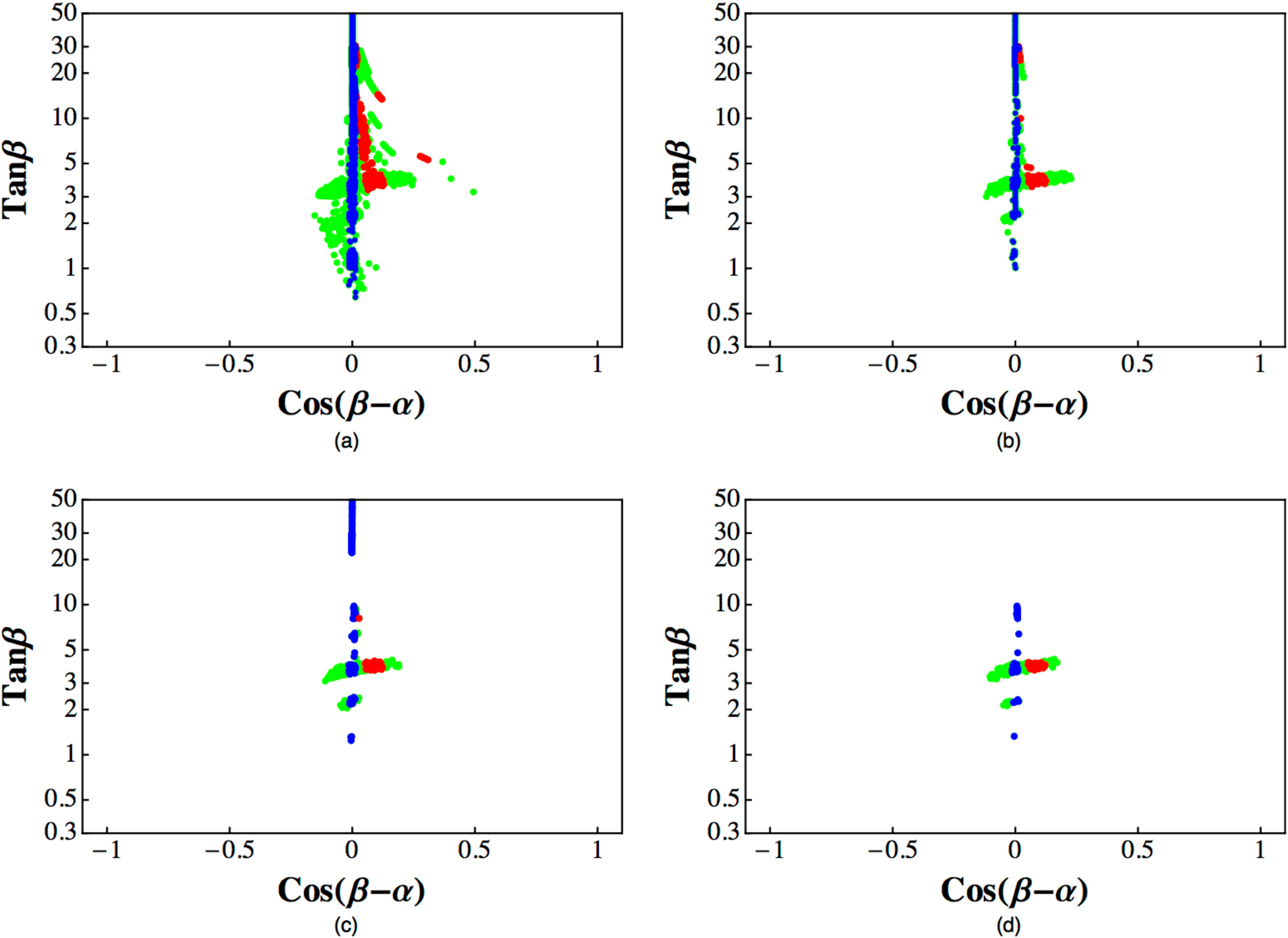, scale=0.397}} \cp{ Plots of  the data points obtained in Fig. \ref{fig4} in the plain ($\cos(\beta-\alpha)$, $\tan\beta$) .
The panels (a), (b), (c) and (d)  correspond to  $\Lambda \simeq 1,~10,~40$  and 100 TeV, respectively.
The points survived all the constraints are displayed in green. Among the survived points,
the ones corresponding to the SM-like Higgs and the ones consistent with the di-photon measurement
at ATLAS are displayed in blue and red, respectively.}\label{fig5}
 \ef
\fig [h] \ct{\ep{figure=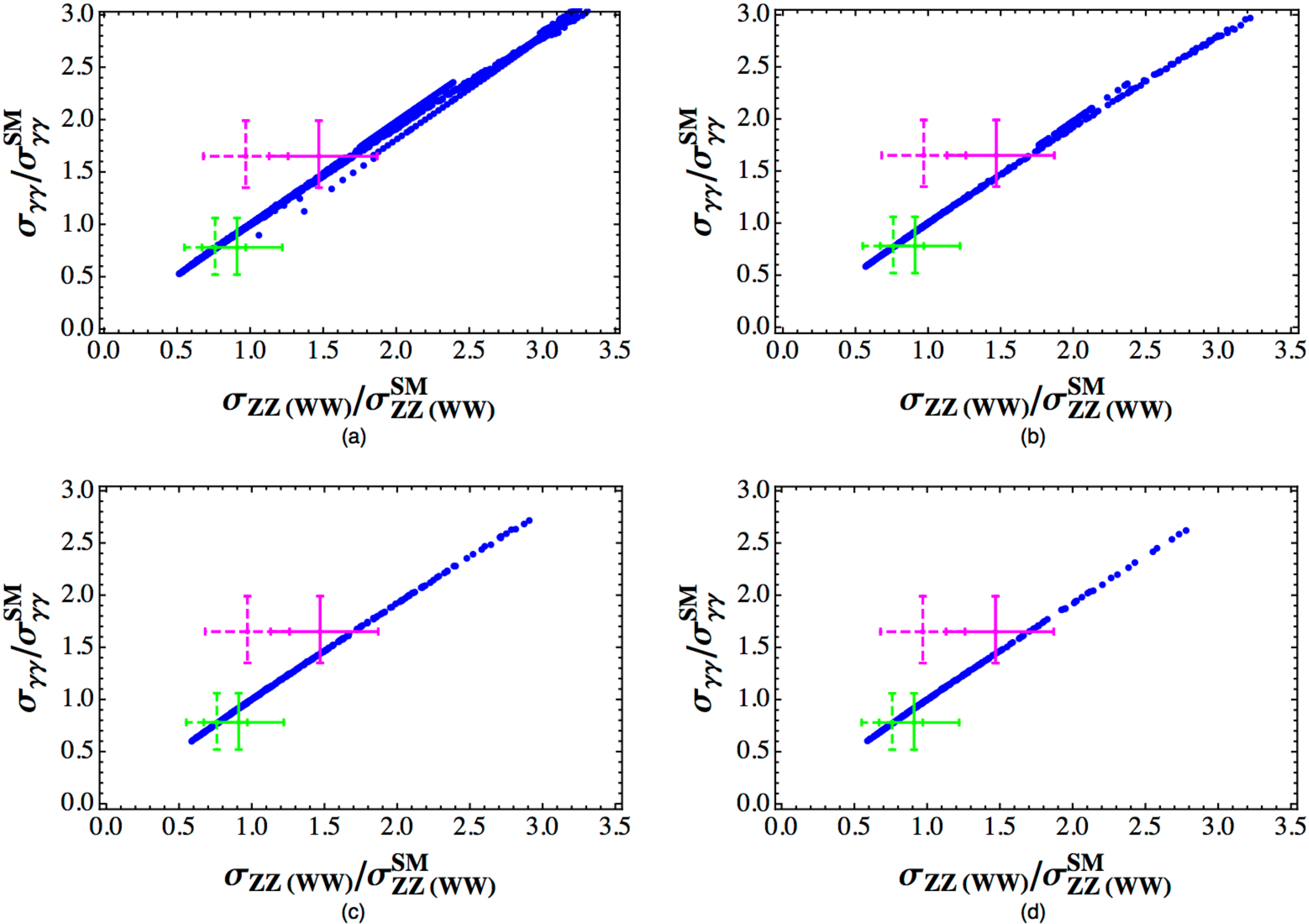, scale=0.5}} \cp{ Predictions of  $\sigma_{\gamma \gamma}/\sigma^{SM}_{\gamma \gamma} ~{\it vs.}~
\sigma_{ZZ^{\ast}(WW^{\ast})}/\sigma_{ZZ^{\ast}(WW^{\ast})}^{SM}$ for  $\Lambda=$ 1 (a), ~10 (b), ~$40$ (c) and 100 (d) TeV.  All blue points correspond to the green ones in Figs. \ref{fig4} and \ref{fig5}. The cross-bars are the same as in Fig. \ref{fig3}.}\label{fig6-1}
 \ef
\fig [h] \ct{\ep{figure=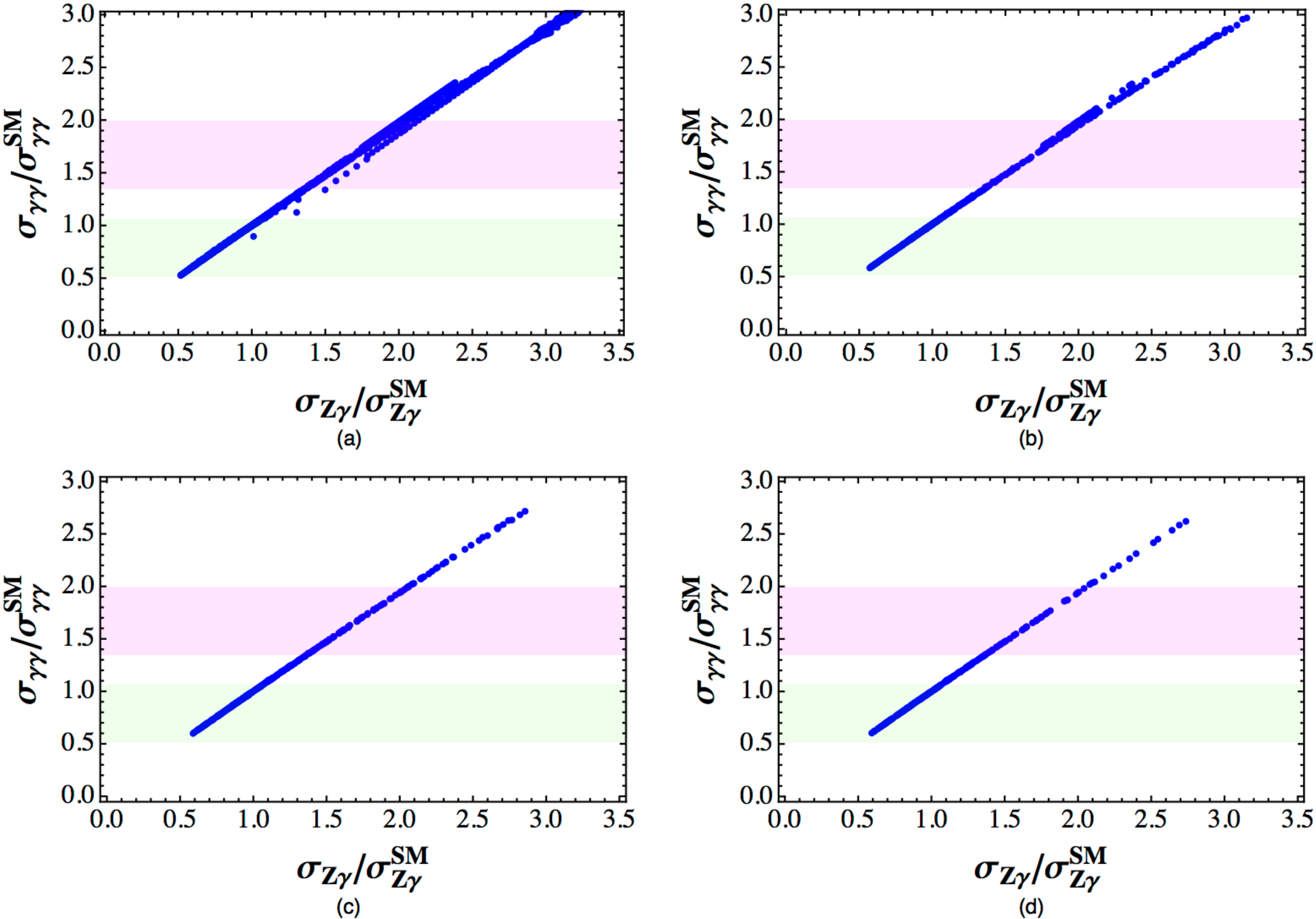, scale=0.5}} \cp{Predictions of  $\sigma_{\gamma \gamma}/\sigma^{SM}_{\gamma \gamma}  {\it vs.} \sigma_{Z\gamma}/\sigma_{Z\gamma}^{SM}$ for the same points in Fig. \ref{fig6-1}  for  $\Lambda=$ 1  (a), ~10  (b), ~$40$  (c) and 100 (d) TeV. The shaded regions are the same as in Fig. \ref{fig3}.}\label{fig6-2}
 \ef
 
In Fig. \ref{fig5}, we plot the allowed data points obtained in Fig. \ref{fig4} in
the plain  ($\cos(\beta-\alpha), \tan\beta$).
The panels (a), (b), (c) and (d) correspond to $\Lambda \simeq 1~\mbox{TeV}, 10 ~ \mbox{TeV}, 40~ \mbox{TeV}$ and 100 TeV, respectively.
The  points survived all the constraints we consider are displayed in green.
Among the points survived all the constraints, the ones corresponding to the SM-like Higgs
with $\cos(\beta-\alpha) \sim 0$  and the ones consistent with the measurement of
the enhanced di-photon at ATLAS  are displayed in blue and red, respectively.
It is likely that the allowed regions get narrowed as the cut off scale increases.
We see that  the region of  $\tan\beta < 0.56 (1.0) $  is excluded in the case of $\Lambda=1 (10~100)$ TeV.

In Fig. \ref{fig6-1}, we show how the predictions of $\sigma_{\gamma \gamma}/\sigma^{SM}_{\gamma \gamma}$ are correlated with those of $ \sigma_{ZZ^{\ast}(WW^{\ast})}/\sigma^{SM}_{ZZ^{\ast}(WW^{\ast})}$ for the allowed regions of parameter space
shown in Fig. \ref{fig4} for $\Lambda=$ 1 (a), ~10 (b), ~$40$ (c) and 100 (d) TeV.
 In Fig. \ref{fig6-2}, we plot the predictions of $\sigma_{\gamma \gamma}/\sigma^{SM}_{\gamma \gamma}~ {\it vs.} ~\sigma_{Z\gamma}/\sigma^{SM}_{Z\gamma}$ for the same parameter space taken in Fig. \ref{fig6-1}. All blue points in Figs. \ref{fig6-1} and \ref{fig6-2} correspond to the green ones
in Fig. \ref{fig4} and \ref{fig5}. The colored cross-bars and shaded regions are the same as in Fig. \ref{fig3}.
As can be seen from Figs. \ref{fig6-1} and \ref{fig6-2},  the allowed region of parameter space is so wide that it could be in consistent with the experimental results of the signal strengths from  not only CMS but also  ATLAS for $\Lambda=1 \sim 100$ TeV.
Contrary to the case of $M_H=126$ GeV,  in this case,
 the coupling of lighter neutral higgs ($h$) to up-type quarks  relative to that of SM, $\frac{\cos\alpha}{\sin\beta}$, can be larger than one, whereas the coupling of $h$ to down type quarks (or charged leptons), $\frac{\sin\alpha}{\cos\beta}$, can be smaller than one, which give rise to enhancements of both (A) and (C) terms in Eq. (\ref{diphotons}).  Those enhancements can be sufficient to enhance the di-photon signal
strength after compensating the possible suppression of the term (B) in Eq.(\ref{diphotons}).

We note that  there are several works in the literature \cite{d2hdm, diphoton, extended diphoton} that study the enhanced di-photon signals in the extended Higgs models, and  the authors in \cite{diphoton} have obtained the parameter region explaining the enhanced di-photon signal in the case of 2HDM, but we have examined the same problem by taking into account the experimental constraints from the LEP experiments and theoretical conditions valid for all renormalization scales up to given cut-off scale. So  we obtain even stronger constraints on $\tan\beta$ and $\alpha$ compared with those obtained in \cite{diphoton}.

\fig [h] \ct{\ep{figure=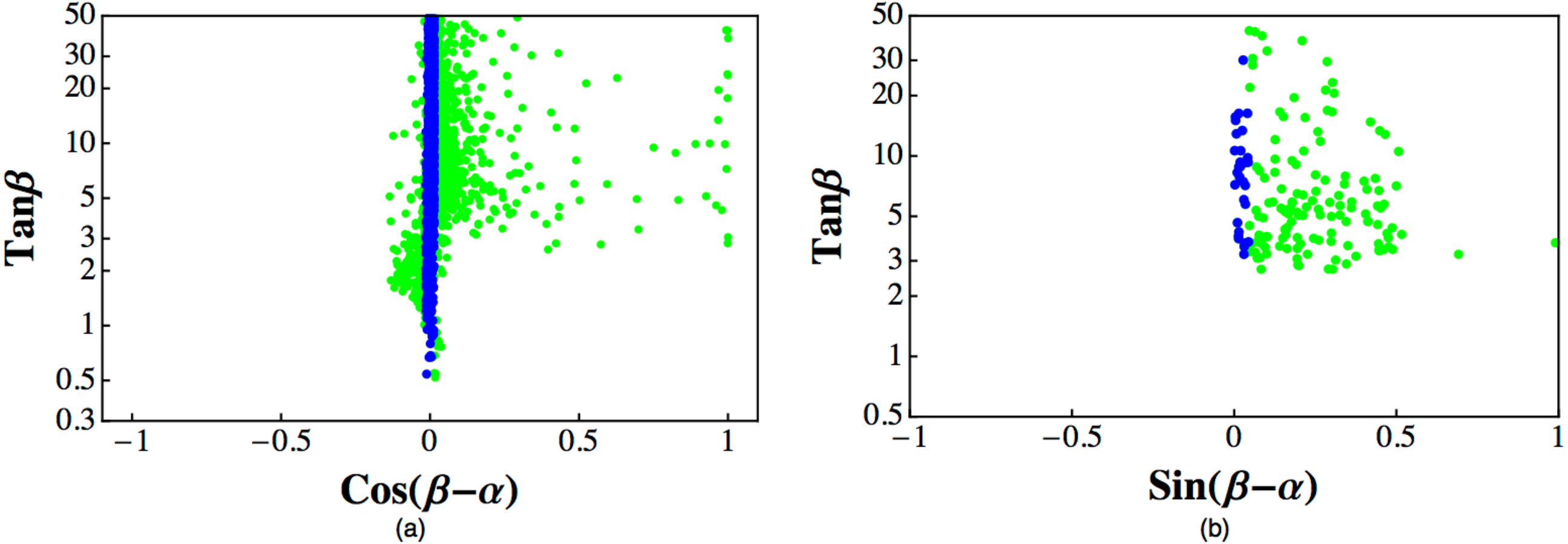, scale=0.5}} \cp{Allowed points in the plains ($\cos(\beta-\alpha), \tan\beta$) (a) and ($\sin(\beta-\alpha), \tan\beta$) (b) for $\Lambda=1$ TeV in type-I 2HDM.  The left (right) panel corresponds to $M_h (M_H) =126$ GeV. The meaning of colors is the same as in Fig. \ref{fig2}.}\label{fig9-1}
\ef
\fig [h] \ct{\ep{figure=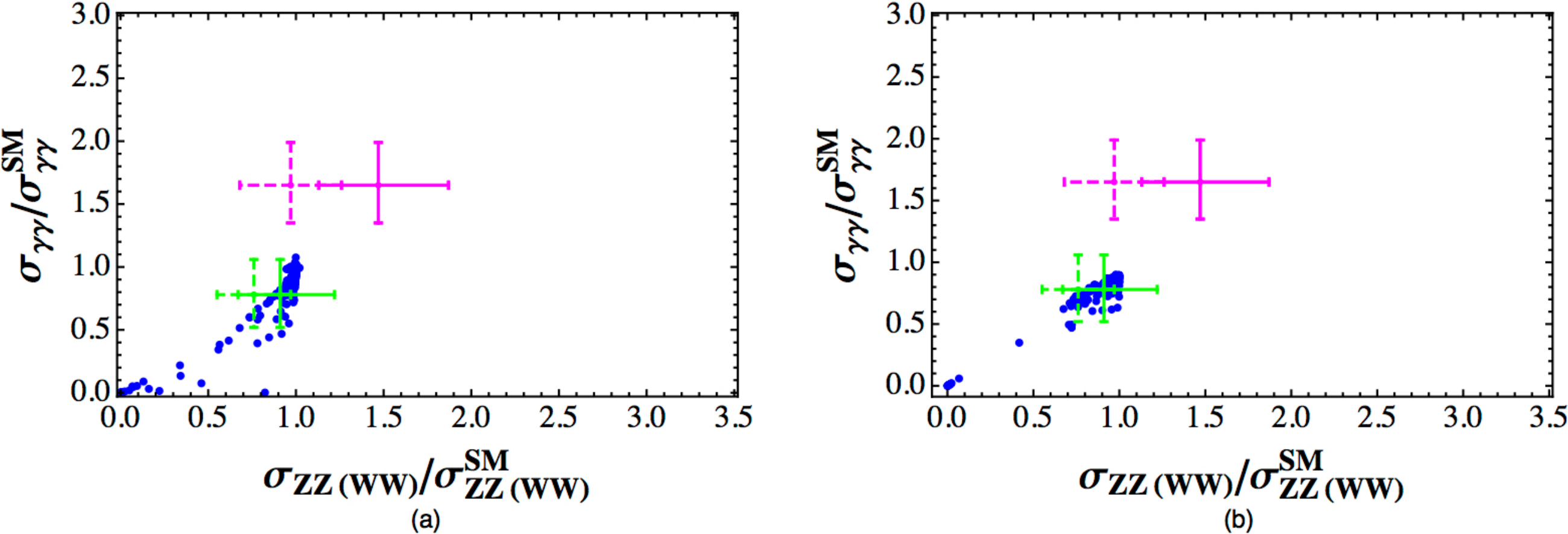, scale=0.5}} \cp { Predictions of  $\sigma_{\gamma \gamma}/\sigma^{SM}_{\gamma \gamma} ~{\it vs.}~
\sigma_{ZZ^{\ast}(WW^{\ast})}/\sigma_{ZZ^{\ast}(WW^{\ast})}^{SM}$ for the same points in Fig. \ref{fig9-1}.
The left (right) panel correspond to $M_h (M_H) =126$ GeV.
The cross-bars are the same as in Fig. \ref{fig3}.}\label{fig9-2}
\ef
\fig [h] \ct{\ep{figure=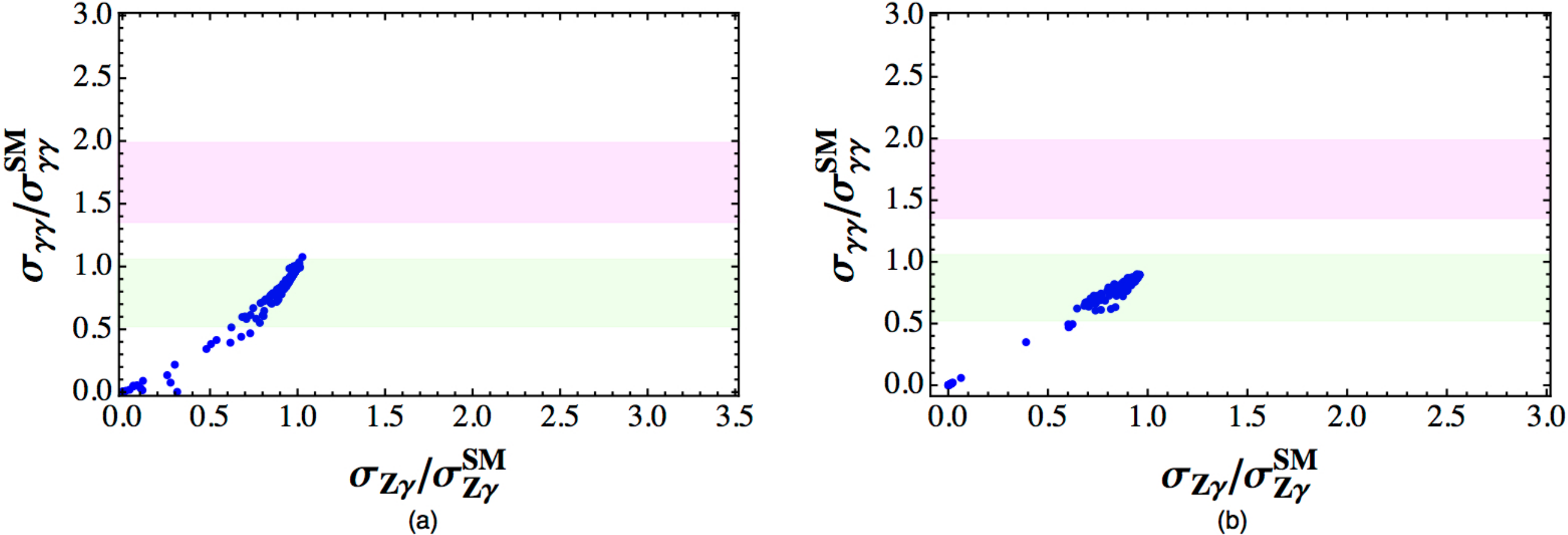, scale=0.5}} \cp{
Predictions of  $\sigma_{\gamma \gamma}/\sigma^{SM}_{\gamma \gamma}  {\it vs.} \sigma_{Z\gamma}/\sigma_{Z\gamma}^{SM}$ for the same points in Fig. \ref{fig9-1}.
 The left (right) panel correspond to $M_h (M_H) =126$ GeV.
The shaded regions are the same as in Fig. \ref{fig3}.}\label{fig9-3}
\ef

Before concluding, remarks on the implications of type-I 2HDM are in order. Compared with type-II 2HDM, the main difference in type-I 2HDM is the Higgs couplings to the fermions. Those couplings are the same as  in the SM but multiplied by  $\frac{\cos\alpha}{\sin\beta}$ and $\frac{\sin\alpha}{\sin\beta}$
for the Higgses $h$ and $H$, respectively. Contrary to type-II model, the Yukawa couplings of down-type quarks can not be enhanced unless $\tan\beta$ is very small. Small values of $\tan\beta$ are excluded or disfavored by perturbativity of Yukawa couplings and constraints from $B$-physics.
In particular, the constraints from $B$-physics lead to different implications of type-I model.
It is known that type-I model is not severely constrained by $b\rightarrow s \gamma$ \cite{Branco}.
Thus, contrary to type-II model, light charged Higgs can be allowed in type-I model, which
can non-negligibly contribute to the Higgs decays and productions.
Thus, the implications of the Higgs signal strengths for the di-photon and $VV^{\ast}$ are different
from those in type-II model. In Fig. \ref{fig9-1}, we display the allowed points by theoretical and experimental constraints in the plains $(\cos(\beta-\alpha), \tan\beta)$  and $(\sin(\beta-\alpha), \tan\beta)$ for
 $\Lambda=1$ TeV in type-I model.
The panels (a) and (b)  correspond to the case of $M_h=126$ GeV and $M_H=126$ GeV, respectively.
Contrary to type-II model, large positive values of $\alpha$ are allowed for the case of $M_h=126$ GeV and  most small values of $\alpha$ are exluded for the case of $M_H=126$ GeV  in type-I model.
For the allowed points, we calculate the signal strengths of the di-photon, gauge boson pairs and $Z\gamma$, and the results are displayed in Figs. \ref{fig9-2} and \ref{fig9-3}.
The left (right) panels correspond to $M_h (M_H) = 126 $ GeV. The colored cross-bars and shaded regions
are the same as in the case of type-II model.
The predictions for both cases are consistent with the recent results from CMS.
While the predictions for the case of $M_h=126$ GeV in type-II model  are so wide that they could cover
the enhancement of di-photon signal observed at ATLAS, those in type-I model do not so.

In conclusion, we have examined the implications of 126 GeV Higgs boson indicated by the recent LHC results  for type II 2HDM.
Identifying the 126 GeV Higgs as either the lighter or heavier of the CP even neutral Higgs, we have obtained the allowed values of Higgs masses and mixing parameters by imposing
 the theoretical conditions and experimental results on the Higgs sectors.
The theoretical conditions taken into account are the vacuum stability, perturbativity and  unitarity required to be satisfied up to a cut-off scale. So, the allowed regions are turned out to be strongly dependent of the cut-off scale.
 We have shown how the experimental constraints on the parameters for Higgs bosons from the LEP
 as well as B physics, and electroweak precision constraints can constrain the parameter spaces further.
Finally, we have found that all the allowed parameter points for the case of  $M_H \sim126$ are incompatible with the enhanced di-photon signal of ATLAS, whereas there exist parameter regions simultaneously accommodating the di-photon and vector boson pair signals observed at the CMS. On the other hand,
in the case of $M_h\sim 126$ GeV, the allowed region of parameter space is so wide that it could be compatible with not only CMS  but also ATLAS experimental results of the signal strengths  for  $\Lambda = 1\sim 100$ TeV.
We have also predicted the signal strengths for $Z\gamma$ channel of the Higgs decay for the allowed parameter regions.

\begin{acknowledgements}
This work was supported by NRF grant funded by MEST (No.2011-0029758).
\end{acknowledgements}

\end{document}